\documentclass[aps,prd,preprint,superscriptaddress]{revtex4}

\usepackage{ulem}
\usepackage{xcolor}
\usepackage[colorlinks=true]{hyperref}

\usepackage{amsmath,amssymb,color,epsfig}
\allowdisplaybreaks[4]

\newcommand{\be}{\begin{equation}}
\newcommand{\ee}{\end{equation}}
\newcommand{\bea}{\setlength\arraycolsep{2pt} \begin{eqnarray}}
\newcommand{\eea}{\end{eqnarray}}
\newcommand{\nn}{\nonumber}
\usepackage{soul}

\begin{document}

\hypersetup{
    linkcolor=blue,
    citecolor=red,
    urlcolor=magenta
}


\title{Neutron stars in Gauss-Bonnet extended Starobinsky gravity}


\author{Zhonghai Liu}
\affiliation{Department of Physics, Synergetic Innovation Center for Quantum Effect and Applications, and Institute of Interdisciplinary Studies, Hunan Normal University, Changsha, 410081, China}

\author{Ziyi Li}
\affiliation{Department of Physics, Synergetic Innovation Center for Quantum Effect and Applications, and Institute of Interdisciplinary Studies, Hunan Normal University, Changsha, 410081, China}

\author{Liang Liang}
\affiliation{Shuda College, Hunan Normal University, Changsha, 410081, China}

\author{Shoulong Li}
\email[Corresponding author: ]{shoulongli@hunnu.edu.cn}
\affiliation{Department of Physics, Synergetic Innovation Center for Quantum Effect and Applications, and Institute of Interdisciplinary Studies, Hunan Normal University, Changsha, 410081, China}

\author{Hongwei Yu}
\email[]{hwyu@hunnu.edu.cn}
\affiliation{Department of Physics, Synergetic Innovation Center for Quantum Effect and Applications, and Institute of Interdisciplinary Studies, Hunan Normal University, Changsha, 410081, China}


\date{\today}

\begin{abstract}

Recently, a class of Gauss-Bonnet extended Starobinsky gravity was proposed, allowing black holes to carry ghost-free massive scalar hair for the first time without requiring additional matter fields. This intriguing feature offers a new perspective for understanding higher-curvature pure gravity and highlights the importance of further studying the potential effects of Gauss-Bonnet extensions in gravitational systems beyond black holes.
In this study, we investigate the properties of neutron stars within this model, focusing on how the higher-curvature terms, particularly the coupling between the Gauss-Bonnet term and the curvature-squared term, impact the stellar structure. We present a detailed analysis of these effects and compute the moment of inertia for rotating neutron stars under the slow-rotation approximation. The substantial differences in the moment of inertia between general relativity and Gauss-Bonnet extended Starobinsky gravity are expected to be detectable through future high-precision observations.

\end{abstract}

\maketitle

\section{Introduction}

Incorporating higher-curvature invariants into general relativity (GR) is a natural generalization of Einstein's theory~\cite{Padmanabhan:2013xyr, Belenchia:2016bvb, Shankaranarayanan:2022wbx}. The simplest example involves a general combination of quadratic curvature polynomials, and the resulting theory has been proven to be renormalizable in four-dimensional spacetimes~\cite{Stelle:1976gc, Stelle:1977ry}, attracting significant and sustained attention. However, the inclusion of higher-order curvature invariants typically leads to a linearized spectrum in maximally symmetric vacua that not only includes the usual massless graviton but also introduces a possible massive scalar and a ghostlike massive spin-2 graviton. While certain special combinations of higher-order curvature polynomials, such as Gauss-Bonnet, more general Lovelock combinations~\cite{Padmanabhan:2013xyr}, and quasi-topological combinations~\cite{Bueno:2016xff, Hennigar:2017ego, Li:2017ncu}, can avoid introducing both massive modes,  it is also notable that some higher-order pure gravity theories introduce only a massive scalar mode, which remains ghost-free. These ghost-free higher-order gravity theories offer intriguing predictions that deviate from GR in regimes where GR alone cannot explain the observed phenomena, without conflicting with the well-established weak-field properties of GR. A prominent example is Starobinsky gravity~\cite{Starobinsky:1980te}, also known as curvature-squared (${\cal R}^2$) gravity, which successfully accounts for cosmic inflation without the need for an inflation field. Numerous variations and extensions of this theory have been widely explored~\cite{Sotiriou:2008rp, DeFelice:2010aj, Nojiri:2010wj}. 
Recently, Liu {\it et al.}~\cite{Liu:2020yqa} considered the coupling of the Gauss-Bonnet combination with Starobinsky gravity and proposed a class of ghost-free higher-order pure gravity theories, called Gauss-Bonnet extended Starobinsky gravity, which allows black holes to carry massive scalar hair. While previous research~\cite{Lu:2015cqa} showed that black holes could support ghostlike massive spin-2 hair without introducing additional matter fields, Gauss-Bonnet extended Starobinsky gravity represents the first model that permits ghost-free massive scalar hair in black holes.  Since this model admits both hairy black holes and Schwarzschild black holes as solutions, current black hole observations are insufficient to place effective constraints on its parameters, which underscores the need for further investigation into the effects of such couplings on other gravitational systems.

Neutron stars, like black holes, serve as crucial astrophysical laboratories for studying the strong-field properties of gravity. Unlike black holes, the structure of neutron stars depends on the coupling between matter and gravity, making it possible to probe, or even rule out, theories that closely resemble GR in vacuum but differ in their treatment of this coupling~\cite{Berti:2015itd, Barack:2018yly, Psaltis:2008bb, Freire:2024adf, Hu:2023vsq, Shao:2022izp, Shao:2022koz, Yagi:2016bkt}. 
Despite the challenges posed by uncertainties in the equation of state (EOS) of neutron stars, EOS-insensitive, nearly universal relations~\cite{Yagi:2016bkt} allow for robust strong-gravity tests without requiring detailed prior knowledge of the EOS. Furthermore, while the effects of gravitational modifications on the stellar structure may vary across different EOS models, these modifications often exhibit common patterns, providing deeper insights into their behavior in strong-field regimes. 
Extensive research has already explored neutron stars within the framework of higher-order curvature pure gravity theories, including Starobinsky gravity~\cite{Cooney:2009rr, Babichev:2009fi, Doneva:2015hsa, Staykov:2015cfa, Ganguly:2013taa, Jaime:2010kn, Arapoglu:2010rz, Orellana:2013gn, Astashenok:2013vza, Yazadjiev:2018xxk, Staykov:2015kwa, Staykov:2014mwa, Capozziello:2015yza, Yazadjiev:2014cza, AparicioResco:2016xcm, Astashenok:2017dpo, Yazadjiev:2015zia, Kobayashi:2008tq, Upadhye:2009kt, Babichev:2009td, Numajiri:2023uif,  Sbisa:2019mae, Pretel:2020rqx, Numajiri:2021nsc, Feola:2019zqg} and its various variants and extensions~\cite{Kobayashi:2008tq,Upadhye:2009kt, Numajiri:2021nsc, Babichev:2009td, Jaime:2010kn, Capozziello:2015yza, Astashenok:2013vza, Numajiri:2023uif, AparicioResco:2016xcm, Bonanno:2021zoy, Santos:2011ye, Astashenok:2021peo, Astashenok:2020cqq, Deliduman:2011nw, Li:2023vbo, Cui:2024nkr, Shamir:2019bcw, Naz:2020mjg, Doneva:2020ped}. These studies encompass both the investigation of neutron star properties and the use of neutron star observables to constrain theoretical parameters. For a comprehensive overview of neutron stars in higher-curvature pure gravity theories, see the reviews in~\cite{Berti:2015itd, Yagi:2016bkt, Stergioulas:2003yp, Paschalidis:2016vmz, Olmo:2019flu, Pani:2011xm} and the references therein.
Given the wide variety of higher-order pure gravity models, their properties differ, leading to varying degrees of influence on the stellar structure. Moreover, due to the different characteristics of these gravitational models, the interpretation of these effects can vary across different frameworks. 
In this study, we investigate neutron stars within the framework of Gauss-Bonnet extended Starobinsky gravity. Our primary objectives are to examine how the coupling between the Gauss-Bonnet term and the curvature-squared term influences the stellar structure and to understand these effects within this specific model. Additionally, we compute the moment of inertia for rotating neutron stars under the slow-rotation approximation~\cite{Hartle:1967he, Hartle:1968si}, as a first step toward exploring nearly universal relations~\cite{Yagi:2016bkt} to test the validity of this model.

The remainder of this paper is organized as follows. In Sec.~\ref{framework}, we provide a review of Gauss-Bonnet extended Starobinsky gravity and derive the modified Tolman-Oppenheimer-Volkoff (TOV) equations for rotating neutron stars under the slow-rotation approximation. The numerical results for the equilibrium configurations and moments of inertia of the neutron stars are presented and analyzed in Sec.~\ref{Results}. In Sec.~\ref{Conclusion}, we summarize our findings and provide concluding remarks.

\section{Gauss-Bonnet extended Starobinsky gravity} \label{framework}

In this section, we first review Gauss-Bonnet extended Starobinsky gravity and its associated equations of motion (EOMs). We then present the modified TOV equations for rotating neutron stars under the slow-rotation approximation within this framework.

\subsection{Equations of motion}

The action $S$ of Gauss-Bonnet extended Starobinsky gravity can be written as
\be
S = \frac{c^4}{16 \pi G} \int{ d^4 x \sqrt{-g} L} +S_\textup{m} \,,
\ee
where $g$, $L$, and $S_\textup{m}$ represent determinants of metric $g_{\mu\nu}$, the gravitational Lagrangian density, and the action for the matter fields, respectively. The symbols $c$ and $G$, representing the speed of light and the gravitational constant, are set to unity ($c=G=1$) in the remainder of the work, as we adopt geometric units. A simple example can be expressed in the following form~\cite{Liu:2020yqa}: 
\be
L = {\cal R} + \frac{\alpha {\cal R}^2}{1 -2 \alpha \beta L_\textup{GB}} \,, \quad \textup{with}  \quad L_\textup{GB} = {\cal R}^2 - 4 {\cal R}_{\mu\nu}{\cal R}^{\mu\nu} +  {\cal R}_{\mu\nu\rho\sigma}{\cal R}^{\mu\nu\rho\sigma} \,, \label{GBES}
\ee
where ${\cal R}$, ${\cal R}_{\mu\nu}$, and ${\cal R}_{\mu\nu\rho\sigma}$ represent Ricci scalar, Ricci tensor, and Riemann tensor, respectively.  Both parameters $\alpha$ and $\beta$ are coupling constants. Setting $\beta = 0$ eliminates the contribution of the Gauss-Bonnet coupling, reducing the theory to the celebrated Starobinsky gravity~\cite{Starobinsky:1980te}. Further setting $\alpha = 0$ returns the theory to standard GR. 

It is worth noting that the construction of such a Lagrangian~(\ref{GBES}), which allows black holes to carry scalar hair, is primarily inspired by the equivalence, in principle, between higher-curvature pure gravity and theories involving additional fields~\cite{Jakubiec:1988ef}, as well as the phenomenon of black holes undergoing spontaneous scalarization  induced by a Gauss-Bonnet coupling~\cite{Doneva:2017bvd, Silva:2017uqg, Antoniou:2017acq}. 
Details of the construction process can be found in Ref.~\cite{Liu:2020yqa} , where the corresponding equivalent Lagrangian can be expressed as
\be
L = {\cal R} +\phi {\cal R} -\frac{1}{2} \mu^2 \phi^2 +U(\phi) L_\textup{GB}\,, \label{edgb}
\ee
with 
\be
U = \frac{1}{2}\beta \phi^2 \,, \quad \mu^2 = \frac{1}{2\alpha} >0 \,,
\ee
where $\phi$ represents a massive scalar field, and $\mu$ can be understood as its mass. The EOMs $E_{\mu\nu}$ and $E_\phi$ are given by
\bea
E_{\mu\nu} &\equiv& (\phi +1) G_{\mu\nu} + g_{\mu\nu} \Box \phi - \nabla_\mu\nabla_\nu \phi + \frac{\mu^2}{4} \phi^2 g_{\mu\nu} + X_{\mu\nu}  = 8\pi T_{\mu\nu} \,, \label{eom1} \\
E_\phi &\equiv& R -\mu^2 \phi + \beta \phi L_\textup{GB} = 0 \,, \label{eom2}
\eea
where $G_{\mu\nu} = {\cal R}_{\mu\nu} - {\cal R} g_{\mu\nu}/2$ is the Einstein tensor, $\nabla_\mu$ is the covariant derivative, $ \Box = \nabla_\mu \nabla^\mu$ is the d'Alembert operator, and $X_{\mu\nu}$ and the energy momentum tensor $T_{\mu\nu} $ are expressed as
\bea
X_{\mu\nu} &=& 8 {\cal R}^\rho{}_{(\mu} \nabla_{\nu)}  \nabla_\rho U - 2 {\cal R} \nabla_\mu \nabla_\nu U  - 4 G_{\mu\nu} \Box U  + 4 {\cal R}^\rho{}_{\mu}{}^\sigma{}_{\nu} \nabla_\rho \nabla_\sigma U - 4 g_{\mu\nu} {\cal R}^{\rho\sigma} \nabla_\rho \nabla_\sigma U \,, \\
T_{\mu\nu} &=& -\frac{2}{\sqrt{-g}} \frac{\partial S_\textup{m}}{\partial g^{\mu\nu}} \,.
\eea
Here, the parentheses denote the symmetrization of the indices enclosed within them.
By solving Eq.~(\ref{eom2}), we obtain 
\be
\phi = \frac{2\alpha {\cal R}}{1- 2\alpha \beta L_\textup{GB}} \,. \label{phicurv}
\ee
Note that the Ricci scalar $\cal R$ is linearly related to scalar field $\phi$ only in Starobinsky gravity, but not in its Gauss-Bonnet extension. Substituting the above equation into the Lagrangian~(\ref{edgb}), it can be verified that the Lagrangian reduces back to Eq.~(\ref{GBES}).

\subsection{Modified Tolman-Oppenheimer-Volkoff equations}

After reviewing Gauss-Bonnet extended Starobinsky gravity, we begin our investigation of neutron stars within this framework. While obtaining solutions for rotating neutron stars with arbitrary angular velocity would provide a more comprehensive understanding of their properties and allow for more accurate calculations of the corresponding observables, solving for rapidly rotating neutron stars in this theory is complex. Moreover, the rotational deformation of even rapidly rotating neutron stars is relatively small~\cite{Haensel:2007yy}, making it more practical to adopt the slow-rotation approximation~\cite{Hartle:1967he, Hartle:1968si} as an initial step. This approximation not only simplifies the analysis but also provides valuable insights into key features such as the frame dragging around neutron stars and their moment of inertia, which makes it widely used in the study of neutron stars in Starobinsky gravity~\cite{Staykov:2014mwa} and other modified gravities~\cite{Berti:2015itd, Yagi:2016bkt, Stergioulas:2003yp, Paschalidis:2016vmz, Olmo:2019flu, Pani:2011xm}. 
 However, for a specific stellar model with a fixed EOS and central density, the range of allowed spin frequencies by the applicability of this approximation may differ. Here, we consider a simplified scenario where the centrifugal force $F$ required to induce deformation in a stellar model with a fixed EOS and central density remains constant, illustrated qualitatively using the relation $F \propto M w^2 R$, where $M, R$, and $w$ represent the mass, radius, and angular velocity of the star, respectively.
If a modified gravity predicts a larger mass and radius for a specific stellar model compared to GR, the applicability of the slow-rotation approximation within this framework requires the maximum angular velocity of the stellar model to be lower than that in GR, and vice versa.

Following the Hartle-Thorne formalism~\cite{Hartle:1967he, Hartle:1968si}, the interior and exterior spacetimes of a slowly and uniformly rotating star with angular velocity $\Omega$ can be described, to first order in  $\Omega$ of the star, by the Lense-Thirring metric ansatz~\cite{Lense:1918zz},
\be
ds^2 = - e^{\lambda(r)} dt^2 + f(r)^{-1} dr^2  +r^2 (d\theta^2 + r^2 \sin^2\theta d\varphi^2) -2\epsilon (\Omega - w(r)) r^2 \sin^2\theta dt d\varphi \,, \label{metric}
\ee
in Schwarzschild coordinates $(t, r, \theta, \varphi)$, where $\epsilon$ is a bookkeeping slow-rotation parameter, while $\lambda, f$, and $w$ are functions of $r$, respectively. The angular velocity $\Omega$ is measured by an observer at rest at some point in the star. The function $w$ represents the angular velocity of the local inertial frame, acquired by an observer falling freely from infinity to the point in the star calculated to first order in $\Omega$, i.e., $\cal O (\epsilon)$. Their difference, $\Omega -w$, represents the angular velocity of the star relative to the local inertial frame.  It is worth noting that the metric~(\ref{metric}) is based on a simple assumption, namely that spacetimes satisfy the circularity condition. For discussions on noncircular cases, please refer to Refs.~\cite{VanAelst:2019kku, Nakashi:2020phm} and references therein.

The slow-rotation approximation implies that the angular velocity $\Omega$ is sufficiently small, so that the changes induced by the rotation remain very slight.  We assume that the influence of the rotation on other metric functions, the scalar field, and the density and pressure of stars is in second order in $\Omega$, and it is, therefore, neglected in this paper. Therefore, the scalar field can be expressed as 
\be
\phi = \phi(r) \,.
\ee
Considering that the star is modeled as a perfect fluid, the energy-momentum tensor can be expressed as
\be
T^{\mu\nu} = (\rho + p) u^\mu u^\nu + p g^{\mu\nu} \,,
\ee
where $u^\mu$ is the four-velocity of the fluid, and is given by
\be
u^\mu = (u^t, 0, 0, \epsilon\Omega u^t) \,,  \quad \textup{with}  \quad u^\mu u_\mu = -1 \,.
\ee
Both pressure $p$ and density $\rho$ are functions of $r$, and their relationship is determined by the EOS
\be
p(r) = P ( \rho(r) ) \,. \label{eos}
\ee

Substituting the above Eqs.~(\ref{metric})---(\ref{eos}) into the EOMs~(\ref{eom1}) and (\ref{eom2}), and solving order-by-order in $\epsilon$, the ${\cal O} (\epsilon^0)$ EOMs correspond to the fluid equilibrium equations, which are the modified TOV equations in Gauss-Bonnet extended Starobinsky gravity. The modified TOV equations are a set of nonlinear ordinary differential equations (ODEs) and can be expressed as follows:
\bea
\lambda^{\prime\prime} &=& F_1 (r, \lambda^{\prime}, f, \phi, \rho) \,, \nn\\
f^{\prime} &=& F_2 (r, \lambda^{\prime}, f, \phi, \rho) \,, \nn\\
\phi^{\prime} &=& F_3 (r, \lambda^{\prime}, f, \phi, \rho) \,, \nn\\
\rho^{\prime} &=& F_4 (r, \lambda^{\prime},  \rho) \,,  \label{zeroorder}
\eea
where $F_i$ was used to simplify the representation of the equations and avoid displaying the specific complex expressions. The detailed formulations are summarized in Appendix~\ref{appendix}.
The ${\cal O} (\epsilon)$ EOM constitutes a second-order linear homogeneous ODE for $w$ and can be expressed as follows:
\be
w^{\prime\prime} + A_1(r) w^\prime  + A_2(r) w = 0 \,.  \label{firstorder}
\ee
where the detailed expressions of $A_1$ and $A_2$  are summarized in Appendix~\ref{appendix}.

We now aim to obtain the specific boundary conditions for Eqs.~(\ref{zeroorder})---(\ref{firstorder}) ensuring  regularity at the center of the star and asymptotic flatness at spatial infinity. Near the center of the star, by considering power expansions for ($\lambda, f, \phi, \rho, w$),  we can express the behavior of the regular solution using a Taylor expansion, as follows:
\bea
\lim_{r \rightarrow 0} \lambda(r) &=& \lambda_0 +\lambda_1 r +\lambda_2 r^2 +{\cal O}(r^3) \,, \nn \\
\lim_{r \rightarrow 0} f (r) &=& f_0 + f_1 r+ f_2 r^2 +{\cal O}(r^3)\,, \nn \\ 
\lim_{r \rightarrow 0} \phi(r) &=& \phi_0 + \phi_1 r + \phi_2 r^2 +{\cal O}(r^3)\,, \nn \\
\lim_{r \rightarrow 0} \rho(r) &=& \rho_0 + \rho_1 r +{\cal O}(r^2)\,, \nn \\
\lim_{r \rightarrow 0}  w(r) &=& w_0 + w_1 r + w_2 r^2 +{\cal O}(r^3) \,, \label{centercond}
\eea
where the nonzero parameters are central density $ \rho_0$, three free parameters ($\lambda_0, \phi _0, w_0$), and $(\lambda_2, f_0, f_2, w_2)$, which are given by
\bea
\lambda_2 &=& (384 \alpha  \beta  \phi _2^2 \phi _0-24 \alpha  \phi _2 \phi _0-24 \alpha  \phi _2+12 \beta  \phi _2 \phi _0^3  +32 \pi  \alpha  \rho_0  (\phi _0+1 )-\phi _0^3-\phi _0^2 \nn\\
&\quad& -96 \pi  \alpha   (\phi _0  (8 \beta  \phi _2-1 )-1 ) P (\rho_0 ))/(24 \alpha   (\phi _0  (8 \beta  \phi _2-1 )-1 ){}^2) \,,\nn \\
f_0 &=& 1 \,, \quad f_2 = \frac{48 \alpha  \phi _2+64 \pi  \alpha  \rho_0+\phi _0^2}{24 \alpha   (\phi _0  (8 \beta  \phi _2-1 )-1 )} \,, \nn \\
w_2 &=& \frac{8 \pi  w_0  (P (\rho _0 )+\rho _0 )}{5  (-8 \beta  \phi _2 \phi _0+\phi _0+1 )} \,,
\eea
and $\phi _2$, which should fulfill the following cubic algebraic equation
\be
B_1 \phi _2^3 +B_2  \phi _2^2 + B_3  \phi _2 + B_4 =0 \,,
\ee
where
\bea
B_1&=& 6144 \alpha  \beta ^2 \phi _0^2  (3 \alpha +\beta  \phi _0^2 )  \,, \nn\\
B_2&=& -2304 \alpha  \beta  \phi _0  (\phi _0+1 )  (3 \alpha +\beta  \phi _0^2 )  \,, \nn\\
B_3&=&  12 (-12 \alpha ^2  (\phi _0+1 )  (\phi _0  (32 \pi  \beta  \rho _0-3 )-3 )+4 \alpha  \beta  \phi _0^2  (\phi _0^2  (3-16 \pi  \beta  \rho _0 )+9 \phi _0+6 )  \nn\\
&\quad& -\beta ^2 \phi _0^6+64 \pi  \alpha  \beta  \phi _0 P (\rho _0 )  (\alpha   (64 \pi  \beta  \rho _0 \phi _0+6 \phi _0+6 )+\beta  \phi _0^3 ))   \,, \nn\\
B_4&=&  (\phi _0+1 ) (-2048 \pi ^2 \alpha ^2 \beta  \rho _0^2 \phi _0  +32 \pi  \alpha  \rho _0  (6 \alpha   (\phi _0+1 )+\beta  \phi _0^3 )-12 \alpha   (\phi _0+1 ) \phi _0 \nn \\
&\quad& +\beta  \phi _0^5-96 \pi  \alpha  P (\rho _0 )  (\alpha   (64 \pi  \beta  \rho _0 \phi _0+6 \phi _0+6 )+\beta  \phi _0^3 )) \,. \label{cubiccondition}
\eea
Setting $\beta = 0$, the above Eq.~(\ref{cubiccondition}) will reduce to a linear equation as the theory reduces to Starobinsky gravity, and $\phi _2$ can be solved as  
\be
\phi _2 = \frac{-16 \pi  \alpha  \rho _0+48 \pi  \alpha  P (\rho _0 )+\phi _0}{36 \alpha } \,,
\ee
For $\beta \ne 0$, $\phi _2$ can be solved as
\be
\phi _2 = -\frac{B_2}{3 B_1} +\frac{2 \sqrt{-C_1} }{\sqrt{3}} \cos \Big[\frac{2 \pi  k}{3}-\frac{1}{3} \cos ^{-1}\Big[\frac{3 \sqrt{3} \sqrt{-\frac{1}{C_1}} C_2}{2 C_1}\Big]\Big] \,, \quad \textup{with} \quad k = 0, 1, 2, \label{phi2sols}
\ee
where
\be
C_1 = -\frac{B_2^2-3 B_1 B_3}{3 B_1^2} \,, \quad C_2 = \frac{2 B_2^3-9 B_1 B_3 B_2+27 B_1^2 B_4}{27 B_1^3} \,.
\ee 
For a given central density $\rho_0$ and a specific EOS, once the appropriate values for ($\lambda_0, \phi_0, w_0$) are chosen, the value of  $\phi_2$ is determined by one of the three solutions from Eq.~(\ref{phi2sols}).  
The specific branch of $\phi_2$ to be selected depends on the particular parameter values and should be determined on a case-by-case basis, ensuring that the solution remains physically meaningful and avoids divergences or inconsistencies in the equations.
Eqs.~(\ref{zeroorder})-(\ref{firstorder}) are then integrated outward from the center of the star until reaching the star’s surface with radius $R$, where the pressure becomes zero, i.e., 
\be
p(R) = 0 \,.
\ee
Then, taking the obtained values for $(\lambda, f, w, \phi)$  at the surface of the star as initial conditions, which are ensured by the continuity condition, we continuously integrate  the vacuum equations outward from the star’s surface to infinity. At asymptotic infinity, the leading falloffs of the ${\cal O} (\epsilon^0)$ metric functions $\lambda, f$, and scalar $\phi$ are given by~\cite{Liu:2020yqa} 
\bea
\lim_{r \rightarrow \infty}  \lambda(r) &=& \ln \big(1 - \frac{2 M}{r} - \frac{\phi_c}{r} e^{-\frac{\mu}{\sqrt{3}}r} \big) \,, \nn \\
\lim_{r \rightarrow \infty}  f(r) &=& 1 - \frac{2 M}{r} +\phi_c \big(\frac{1}{r} + \frac{\mu}{\sqrt{3}} \big) e^{-\frac{\mu}{\sqrt{3}}r} \,, \nn \\
\lim_{r \rightarrow \infty}  \phi(r) &=&  \frac{\phi_c}{r} e^{-\frac{\mu}{\sqrt{3}}r} \,. \label{inftycond0}
\eea
In the above equations, the Yukawa falloff term arises from taking the weak-field limit of the theory, and $M$ and $\phi_c$ represent the mass and massive scalar charge of the neutron star, respectively. 
Generally, in higher-curvature pure gravity, the linear spectrum may include not only the usual massless graviton but also a massive scalar mode and a ghostlike massive spin-2 mode~\cite{Stelle:1976gc, Stelle:1977ry}. Both Starobinsky gravity and its Gauss-Bonnet extension are ghost-free theories, and their linear spectrum contains only a massive scalar mode, allowing static spherically symmetric vacuum solutions to carry a long-range ghost-free scalar charge. The key distinction between these two theories is that when the gravitational source is a black hole, the scalar charge must be zero in Starobinsky gravity, whereas in its Gauss-Bonnet extension, the scalar charge can be nonzero~\cite{Liu:2020yqa}. In certain higher-curvature polynomial gravity models, such as the pure Gauss-Bonnet combination, more general Lovelock combinations~\cite{Padmanabhan:2013xyr, Belenchia:2016bvb, Shankaranarayanan:2022wbx}, or quasi-topological combinations~\cite{Bueno:2016xff, Hennigar:2017ego, Li:2017ncu}, the linearized gravity in four-dimensional spacetimes is identical to GR, and their static spherically symmetric vacuum solutions cannot acquire any additional long-range hair~\cite{Li:2023vbo}, regardless of whether the source is a black hole or a star. The effects of the Yukawa term introduced by Starobinsky gravity on the structure of neutron stars have been studied extensively in the literature, such as in Refs.~\cite{Sbisa:2019mae, Bonanno:2021zoy}. In the next section, we will further explore the additional effects introduced by the Gauss-Bonnet coupling.

The asymptotic behavior of angular velocity $w$ at infinity can be written as
\be
\lim_{r \rightarrow \infty}  w(r) = \Omega - \frac{2 J}{r^3} \,,\label{inftycond1}
\ee
where $J$ represents the angular momentum of the neutron star.
The moment of inertia $I$ of the star is defined as
\be
I = \frac{J}{\Omega} \,. \label{moi}
\ee 
At the end of this section, we adopt  the symbol ``$\star$'' as a subscript to represent the dimension of various physical quantities, 
\be
M_\odot  \sim r_\star \sim \phi_{\star} \sim {\cal R}_\star^{-\frac12} \sim p_\star^{-\frac12} \sim \rho_\star^{-\frac12} \sim \alpha_\star^{\frac12} \sim \mu_\star^{-1} \sim \beta_\star^{\frac12} \sim I_\star^{\frac13} \,, \label{dimension}
\ee
with the solar mass $M_\odot$ serving as the reference unit. These quantities allow us to transform the Eqs.~(\ref{zeroorder}) and (\ref{firstorder}) to a dimensionless form.

\section{Equilibrium configurations and moments of inertia}  \label{Results}

Having established all the equations to be solved, namely Eqs.(\ref{zeroorder})–(\ref{firstorder}), along with the boundary conditions at the center(\ref{centercond}) and at asymptotic infinity~(\ref{inftycond0})–(\ref{inftycond1}), we now specify the EOS for the neutron stars. In this paper, we use two realistic EOSs: SLy~\cite{Douchin:2001sv}, with its analytical form provided in Appendix~\ref{appendixB} based on~\cite{Haensel:2004nu}, and AP4~\cite{Akmal:1997ft,Lattimer:2000nx}. To solve the set of nonlinear ODEs~(\ref{zeroorder})–(\ref{firstorder}), we treat it as a boundary value problem and apply the shooting method. 
We match all the functions $(\lambda, \lambda^\prime, f, \phi, w, w')$ to be solved at a distance of 10 times the stellar radius, ensuring that the error does not exceed $10^{-6}$.
In the following sections, we will analyze the results for both the stellar structures under hydrostatic equilibrium and the moments of inertia of stars in the context of first-order slow rotation.

\subsection{Equilibrium configurations}

\begin{figure}[]
\includegraphics[width=0.95\linewidth]{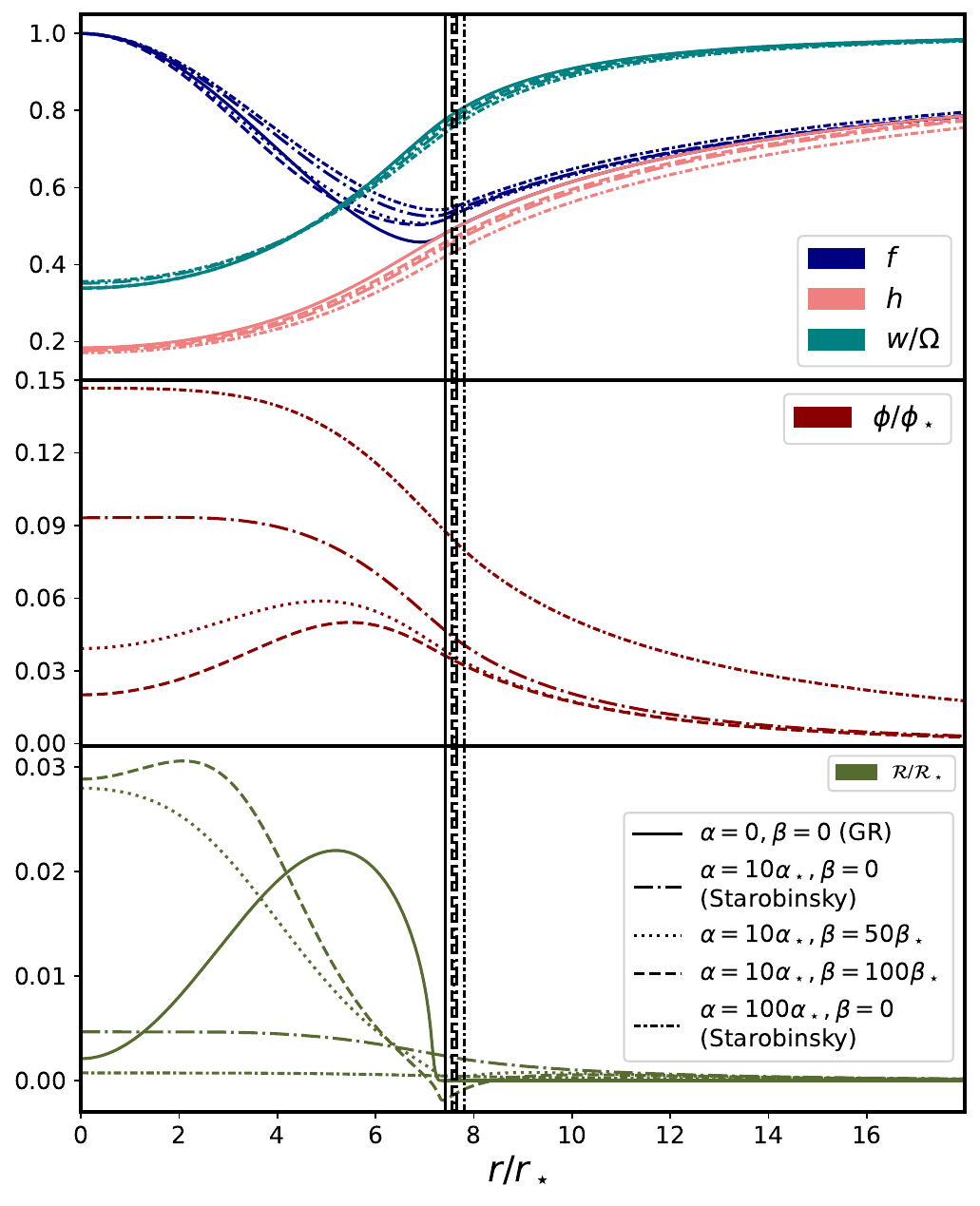}
\caption{\label{fig:0thsolution}The  plots illustrate the numerical solutions of the functions ($h, f, w$) (upper), scalar field $\phi$ (middle), and Ricci scalar $\cal R$ (bottom) in the framework of Gauss-Bonnet extended Starobinsky gravity with coupling constants (1) $\alpha =0 \,, \beta =0 $ , (2) $\alpha =10 \alpha_\star \,, \beta =0 $ , (3) $ \alpha =10 \alpha_\star \,, \beta =50 \beta_\star$ , (4) $\alpha =10 \alpha_\star \,, \beta =100 \beta_\star$, and (5) $\alpha =100 \alpha_\star \,, \beta =0 $ , respectively. The curves represent the solutions and the vertical lines indicate the positions of the radii of the neutron stars. The central density is set to $\rho_0 = 1.7 \times 10^{15} \textup{g/cm}^3$, and the EOS used is SLy.}
\end{figure}

To illustrate the effect of the coupling parameters $\alpha$ and $\beta$ on stellar structures of neutron stars, we first plot the metric functions $h = e^\lambda, f$, normalized angular velocity $w/\Omega$, scalar field $\phi$, and Ricci scalar $\cal R$ as functions of $r$ for several cases with different parameter values in Fig.~\ref{fig:0thsolution}, where the EOS used is SLy and the central density is set to a relatively high value, $\rho_0 = 1.7 \times 10^{15} \textup{g/cm}^3$.
From the upper and middle panels of Fig.~\ref{fig:0thsolution}, the most notable features are that the Birkhoff's theorem is violated in higher-order gravity ($\alpha \ne 0$) and the massive scalar charge of the star can be nonzero. For $\alpha=0$, i.e., in the framework of GR, the metric functions $h$ and $f$ typically have $h \ne f$ near the center, and  gradually converge to become equal  as the radius $r$ increases before reaching the surface. This smooth convergence connects them to the Schwarzschild metric ($h_{\textup{Sch}} = f_{\textup{Sch}} = 1 - 2 M/r$), ensuring that the external vacuum region is uniquely described by the Schwarzschild solution. For $\alpha \ne 0$,  regardless of whether $\beta=0$, the external vacuum region of the star exhibits $h \ne f$. However, changes in the value of $\beta$ will also affect the metric functions and scalar charges of the star for a given model. By comparing cases (1) $\alpha=0$, (2) $\alpha=10$, and (5) $\alpha=100$ with $\beta=0$ fixed, the effect of increasing $\alpha$ on the metric functions becomes clear. Similarly, comparing cases (2) $\beta=0$, (3) $\beta=50$, and (4) $\beta=100$ with $\alpha=10$ fixed highlights the impact of increasing $\beta$. Notably, the curvature-squared term and the Gauss-Bonnet coupling produce opposite effects.

The changes in the Ricci scalar, as shown in the bottom panel of Fig.~\ref{fig:0thsolution}, provide an alternative perspective by allowing us to directly understand the effects of the higher-curvature terms on the equilibrium configurations from the standpoint of pure gravity(\ref{GBES}), rather than the equivalent scalar-tensor formulation~(\ref{edgb}). 
From the bottom panel of Fig.~\ref{fig:0thsolution}, the scalar curvature does not decrease to zero outside the star as it does in GR. In Starobinsky gravity, this phenomenon was understood as the emergence of the so-called ``gravitational sphere'' around the star, as discussed  in works, such as~\cite{Astashenok:2017dpo}. However, in the same framework, static spherically symmetric black holes are uniquely described by the Schwarzschild metric, where the curvature outside the black hole is zero. Thus, under this interpretation, the so-called gravitational sphere appears only around stars, not black holes.
If the absence of a gravitational sphere around black holes is attributed to structural differences between stars and black holes, then in the context of Gauss-Bonnet extended Starobinsky gravity, which allows for both Schwarzschild black holes and hairy black holes, the physical interpretation of the gravitational sphere leads to inconsistencies. Specifically, it does not adequately explain why the gravitational sphere would appear around stars and hairy black holes, but not around Schwarzschild black holes, in this broader ghost-free, higher-curvature gravity theory.
In our view, the nonzero vacuum curvature near the outer boundary of these gravitational sources in higher-order gravity is simply a consequence of the violation of Birkhoff's theorem. In GR, the geometry of the vacuum is uniquely described by the Schwarzschild metric, meaning that the scalar curvature vanishes outside static, spherically symmetric gravitational sources, whether they are black holes or stars. However, in higher-curvature gravity, Birkhoff's theorem no longer holds, allowing the external vacuum solutions for different gravitational sources to deviate from the Schwarzschild metric, irrespective of the nature of the source. It is important to note that this nonzero scalar curvature exists only near the boundary of compact gravitational sources, i.e., in the strong-field regions. In the weak-field regions, especially at asymptotic infinity, the scalar curvature vanishes as both $h$ and $f$ approach the Schwarzschild metric in the weak-field limit.

The idea proposed in Ref.~\cite{Astashenok:2017dpo}, suggesting that the gravitational sphere provides an additional contribution to the gravitational mass, can also be naturally explained by the violation of Birkhoff's theorem, without the need to invoke the concept of the gravitational sphere.  Since the geometric description outside the star in higher-curvature gravities deviates from the Schwarzschild metric, if, from the outset, $f(r)$ in the metric ansatz~(\ref{metric}) is replaced by $1-2 M(r)/r$ to represent $g_{rr}$, then it follows naturally that the mass function $M(r)$ at the stellar surface is not equal to a constant.

\begin{figure}[]
\includegraphics[width=0.95\linewidth]{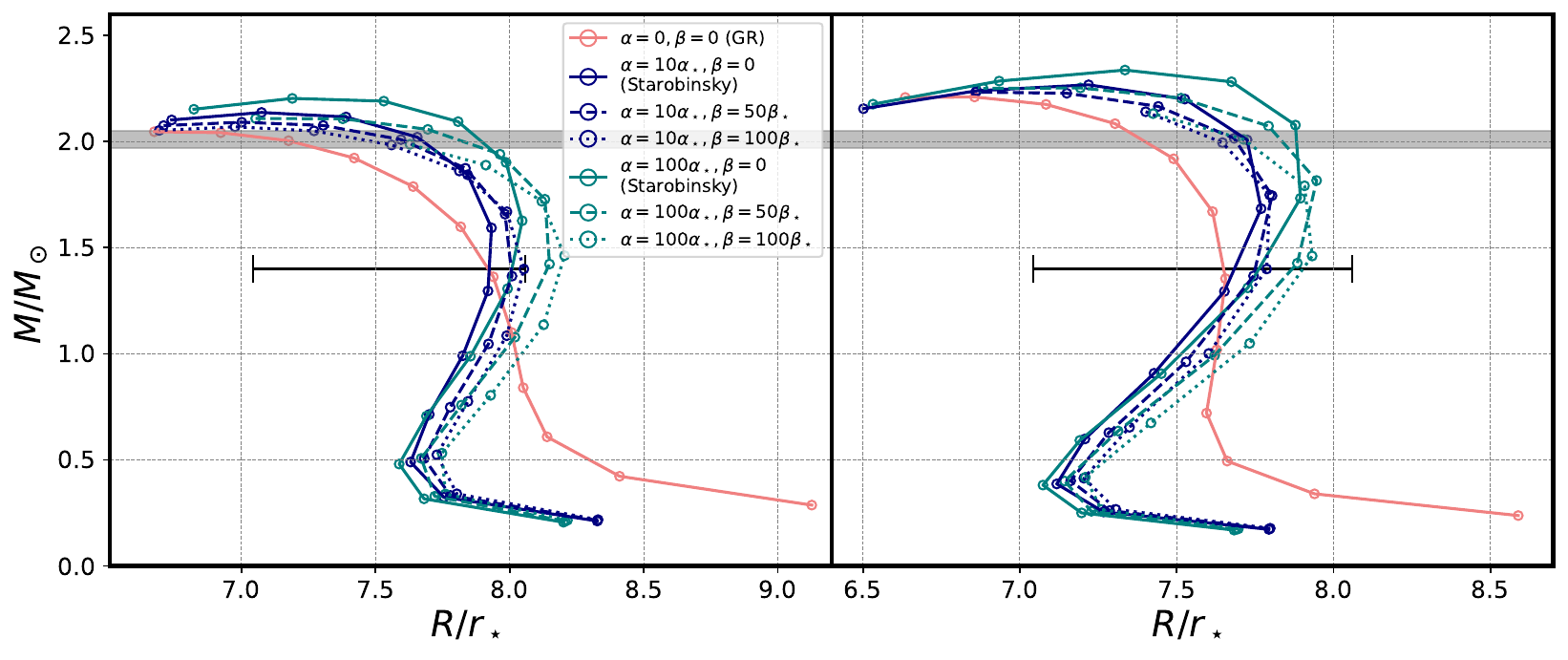}
\caption{\label{fig:MRrelation}The  plots illustrate the mass-radius ($M-R$) relations for EOSs SLy (left) and AP4 (right) in the framework of Gauss-Bonnet extended Starobinsky gravity. Different color and line styles represent different coupling constants $\alpha$ and $\beta$. The grey shaded region represents a neutron star with a mass of $2.01 \pm 0.04 M_\odot$~\cite{Antoniadis:2013pzd}. The error bar (black line) represents  a neutron star with mass $1.4 M_{\odot}$ and radius $R = 7.45^{+0.61}_{-0.41} r_\star$~\cite{Capano:2019eae}. }
\end{figure}

Next, we plot the mass-radius relations for the SLy and AP4 EOSs over a range of central densities $\rho_0$ for several cases with different values of the parameters $\alpha$ and $\beta$ in Fig.~\ref{fig:MRrelation}, to provide a more comprehensive understanding of the effects of higher-curvature terms on different central density configurations.
According to Ref.~\cite{Liu:2020yqa}, larger values of $\alpha$ and $\beta$ allow for hairy black holes with a wider range of compactness.  Therefore, compared to the parameter settings in Fig.~\ref{fig:0thsolution}, we have added cases with $\beta=50$ and $\beta=100$ for $\alpha=100$ to further illustrate.
From Fig.~\ref{fig:MRrelation},  although the EOSs differ, leading to variations in the mass-radius relations, both still exhibit some common features.
As $\alpha$ increases, the mass and radius of neutron stars with higher central densities gradually increase, while those with lower central densities gradually decrease~\cite{Yazadjiev:2014cza}. 
The phenomenon observed in Fig.~\ref{fig:0thsolution}, where the Gauss-Bonnet coupling and the curvature-squared term produce opposite effects in high central density regions, also persists in low central density regions. Specifically, fixing $\alpha$, as $\beta$ increases, the mass and radius of neutron stars with higher central densities gradually decrease, whereas those with lower central densities gradually increase. This indicates that the presence of the Gauss-Bonnet coupling reduces the maximum mass of neutron stars. 
The opposite effects exhibited by the Gauss-Bonnet coupling and the curvature-squared term suggest a broad viable parameter space for $\alpha$ and $\beta$. Even though larger values of $\alpha$ might introduce corrections to macroscopic observables, such as the mass and radius of neutron stars, which could potentially exceed current observational constraints, the effects of $\beta$ can counterbalance these corrections, thereby maintaining consistency with observations.
 However, similar to hairy black holes in this theory, not all regions of parameter space permit the existence of neutron stars. As the parameter $\beta$ increases for a fixed $\alpha$, the central densities that allow for neutron star solutions decrease. From the perspective of the equivalent scalar-tensor theory, both of these features are also observed in the mass-radius relations of neutron stars in other types of Einstein-dilaton Gauss-Bonnet gravities~\cite{Pani:2011xm, Doneva:2017duq, Xu:2021kfh}.
Thirdly, although $\alpha$ and $\beta$ have the same dimensions according to Eq.~(\ref{dimension}), the effects of $\beta$ on the mass and radius are significantly smaller than those of $\alpha$, indicating that the curvature-squared terms in this theory remain dominant. 
Fourth, as an initial step to constrain the theoretical parameters $\alpha$ and $\beta$ using neutron star observations, we also include in Fig.~\ref{fig:MRrelation} the data for a $2.01\pm 0.04 M_{\odot}$ neutron star in a neutron star-white dwarf binary system~\cite{Antoniadis:2013pzd} and the radius constraints of $7.45^{+0.61}_{-0.41} r_\star$ for a $1.4 M_{\odot}$ neutron star, derived from multimessenger observations and nuclear physics~\cite{Capano:2019eae}. Figure~\ref{fig:MRrelation} shows that the chosen parameter values are capable of predicting neutron stars with masses of approximately $2 M_{\odot}$ for both the SLy and AP4 EOSs. For a $1.4 M_{\odot}$ neutron star, however, the parameter sets ($\alpha =100, \beta =50$) and ($\alpha =100, \beta =100$) deviate from observational constraints for the SLy EOS but remain consistent with the constraints for the AP4 EOS.
 Finally, we would like to briefly discuss additional factors that influence the mass and radius of neutron stars. In our study, we focus on neutron stars in the slow-rotation limit, where centrifugal deformations are negligible, thus allowing us to disregard the effects of rotation on the mass of a given stellar model. However, for rapidly rotating neutron stars, centrifugal deformation can lead to an increase in both mass and radius. This effect has been discussed in the contexts of both GR~\cite{Cook:1993qr} and Starobinsky gravity~\cite{Yazadjiev:2015zia}. Furthermore, the maximum stable mass of a neutron star may increase if a strong magnetic field is generated in the stellar core~\cite{Suvorov:2021ymy} or if high temperatures within the star provide additional thermal pressure~\cite{Kaplan:2013wra}. These interesting properties also warrant further exploration within the framework of Gauss-Bonnet extended Starobinsky gravity.

\begin{figure}[]
\includegraphics[width=0.95\linewidth]{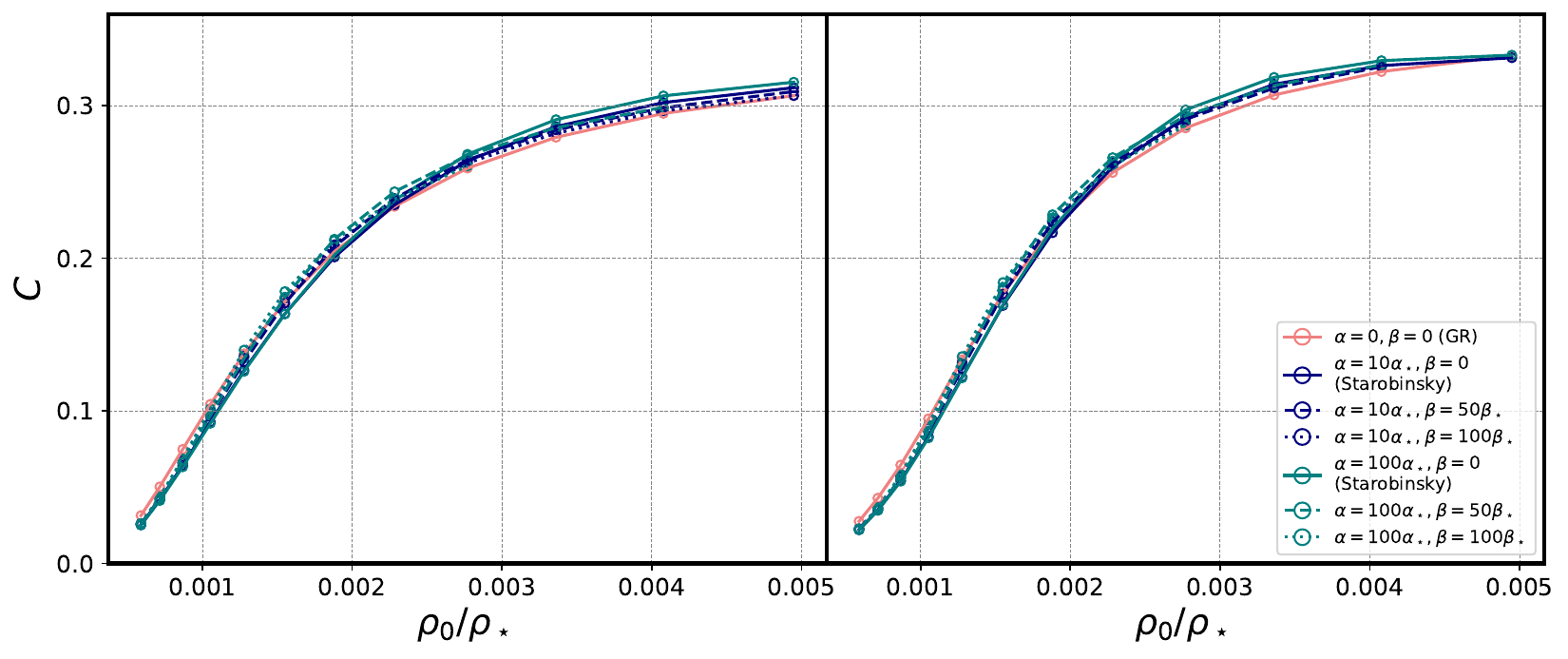}
\caption{\label{fig:rhoCrelation}The plots illustrate the compactness--central density ($C-\rho_0$) relations for EOSs SLy (left) and AP4 (right) in the framework of Gauss-Bonnet extended Starobinsky gravity. Different color and line styles represent different coupling constants $\alpha$ and $\beta$.}
\end{figure}

After studying the mass-radius relations, we further investigate whether these changes in mass $M$ and radius $R$ lead to significant variations in neutron star compactness $C = M/R$. The motivation for this analysis stems from a recent finding~\cite{Li:2023vbo} that a class of higher-curvature pure gravity theories, known as quasi-topological gravity, permits the existence of stars more compact than black holes. This discovery prompted us to explore whether similar effects can be observed in Starobinsky gravity and its Gauss-Bonnet extension, which also belong to the realm of higher-curvature pure gravity.
Both Starobinsky gravity and quasi-topological gravity are ghost-free theories and share a key feature: while static spherically symmetric vacuum solutions are not unique, the static spherically symmetric black hole solution is uniquely described by the Schwarzschild metric. 
If the curvature-squared term in these theories can generate a significant repulsive force, weakening gravitational attraction, neutron stars could potentially reach a compactness greater than that of black holes.
For Gauss-Bonnet extended Starobinsky gravity, although it allows for the existence of hairy black holes that are more compact than Schwarzschild black holes, it is still necessary to explore how the compactness of neutron stars changes within this theory. Therefore, we present the relationship between compactness $C$ and central density $\rho_0$ in Fig.~\ref{fig:rhoCrelation}. 
We find that, for a given EOS and central density $\rho_0$, unlike the quasi-topological combination case~\cite{Li:2023vbo}, while the curvature-squared term and Gauss-Bonnet coupling affect the mass-radius relation, their impact on compactness is negligible.

\subsection{Moments of inertia}

\begin{figure}[]
\includegraphics[width=0.95\linewidth]{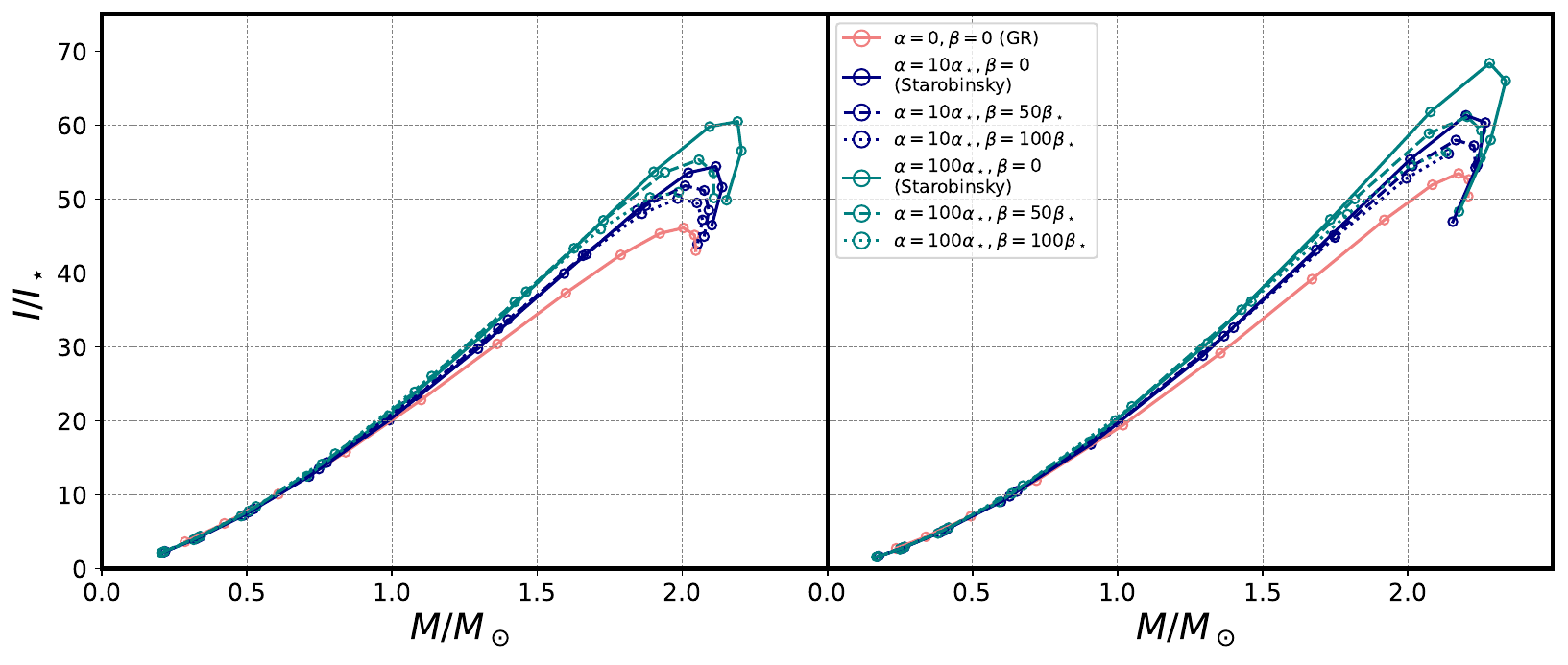}
\caption{\label{fig:IMrelation}The plots illustrate the moment of inertia--mass ($I-M$) relations for EOSs SLy (left) and AP4 (right) in the framework of Gauss-Bonnet extended Starobinsky gravity. Different color and line styles represent different coupling constants $\alpha$ and $\beta$.}
\end{figure}

\begin{figure}[]
\includegraphics[width=0.95\linewidth]{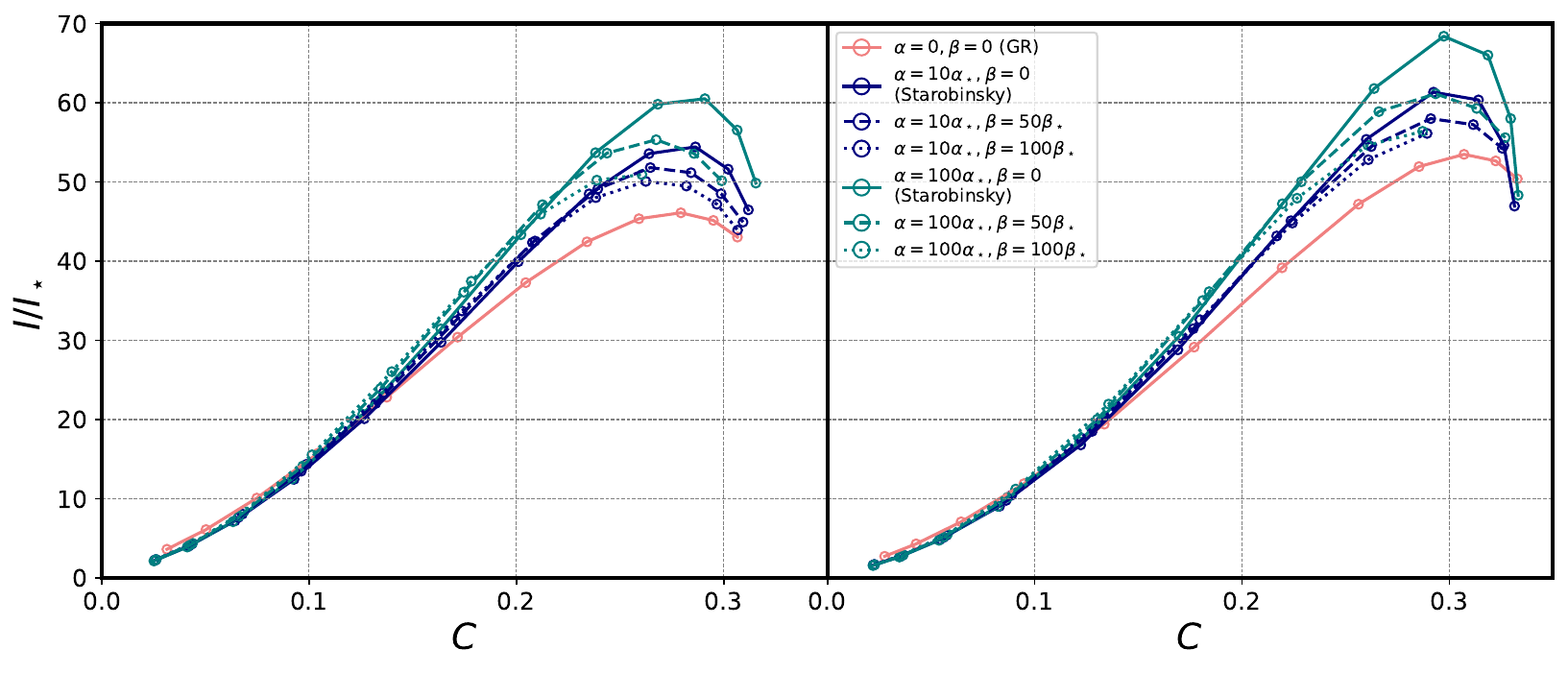}
\caption{\label{fig:ICrelation}The  plots illustrate the moment of inertia--compactness $I-C$ relations for EOSs SLy (left) and AP4 (right) in the framework of Gauss-Bonnet extended Starobinsky gravity. Different color and line styles represent different coupling constants $\alpha$ and $\beta$. }
\end{figure}

As mentioned earlier, one macroscopic observable of neutron stars that can be obtained using the slow-rotation approximation is the moment of inertia $I$, as given by Eq.~(\ref{moi}). We present the relationships between the moment of inertia of neutron stars and mass ($I-M$ relation) in Fig.~\ref{fig:IMrelation}, and the relationships between the moment of inertia and compactness ($I-C$ relation) in Fig.~\ref{fig:ICrelation}, for the EOSs SLy and AP4, taking into account different values of the coupling constants $\alpha$ and $\beta$.
We find that the $I-M$ and $I-C$ relations for different EOSs (SLy and AP4) exhibit similar properties, so we will focus on explaining these commonalities.
Similar to GR, in higher-curvature gravity, both the moment of inertia and mass of neutron stars reach a maximum, but the moment of inertia typically peaks before the star reaches its maximum mass. For neutron stars with the same mass, the effects of higher-curvature terms on the moment of inertia are more pronounced at higher masses, while at lower masses, the impact is weaker and can even be negligible~\cite{Staykov:2014mwa}.
Compared to the curvature-squared term, the Gauss-Bonnet coupling requires a larger mass to produce a significant effect on the moment of inertia, and the effect is opposite to that produced by the curvature-squared term. Nonetheless, the significant differences in the moment of inertia between GR and Gauss-Bonnet extended Starobinsky gravity for various coupling parameters have the potential to be precisely distinguished by future high-precision pulsar timing techniques~\cite{Freire:2024adf, Hu:2020ubl, Miao:2021gmf, Shao:2014wja, Weltman:2018zrl}, which could help constrain the values of these coupling parameters.

\section{Conclusion} \label{Conclusion}

In this work, we investigated neutron stars within the framework of Gauss-Bonnet extended Starobinsky gravity. First, we derived the modified TOV equations and obtained numerical solutions using the shooting method with two realistic EOSs, SLy and AP4. We then examined the effects of higher-curvature terms, particularly the coupling between the Gauss-Bonnet term and the curvature-squared term, on the stellar structure, and provided explanations within the model's framework. Finally, we calculated the moment of inertia of neutron stars and analyzed its relationship with mass and compactness.

Similar to Starobinsky gravity, the Gauss-Bonnet extension also violates Birkhoff's theorem, resulting in the static spherically symmetric vacuum solution no longer being uniquely described by the Schwarzschild metric. Consequently, the scalar curvature near the surface of a star remains nonzero, while it vanishes outside a Schwarzschild black hole. In Starobinsky gravity, this has been interpreted as the emergence of a gravitational sphere around stars but not black holes. However, in Gauss-Bonnet extended Starobinsky gravity, which allows both Schwarzschild and hairy black holes, the scalar curvature outside the horizon of a hairy black hole is also nonzero. Thus, the concept of a gravitational sphere fails to consistently explain this phenomenon, and we instead interpreted it directly as a result of Birkhoff's theorem being violated in these models.

Additionally, we found that compared to the curvature-squared term, the Gauss-Bonnet coupling generally had opposite and less dominant effects on observables, such as mass, radius, and moment of inertia. Moreover, both terms had a negligible impact on the star's compactness. However, the significant differences in moment of inertia between GR and Gauss-Bonnet extended Starobinsky gravity could be precisely distinguished by future high-precision pulsar timing techniques, offering a way to constrain the coupling parameters. This will provide valuable insights into the nature of higher-order pure gravity.

Lastly, we emphasize that studying the mass-radius relationship and the moment of inertia of rotating neutron stars under the slow-rotation approximation is only the first step in testing Gauss-Bonnet extended Starobinsky gravity.
To impose tighter constraints on the theoretical parameters, it is essential to investigate additional observables, such as the quadrupole moment and tidal Love numbers, which can leverage EOS-insensitive nearly universal relations, such as the I-Love-Q relations~\cite{Yagi:2016bkt, Yagi:2013bca, Yagi:2013awa}, between different observables. 
There are still many other aspects related to neutron stars within this gravitational framework that warrant further investigation, such as the case of rapidly rotating stars, additional observables, stability under adiabatic perturbations, quasinormal modes~\cite{Kokkotas:1999bd}, and so on.

\appendix 

\section{Analytic expressions of modified TOV equations} \label{appendix}

For the sake of completeness, we present the full form of the TOV Eqs.~(\ref{zeroorder})---(\ref{firstorder}) here. The Eq.~(\ref{zeroorder}) is given by
\bea
\lambda'' &=& F_1 (r, \lambda^{\prime}, f^{\prime}, \phi, \rho) \nn\\
&\equiv& \Big(-4096 \pi ^2 \alpha ^2 \beta  P(\rho )^2  (f  (r \lambda'-2 )+2 ) r^5+\beta  \phi ^4  (-5 f  (r \lambda'-2 )-2 ) r^5 \nn\\
&\quad&-16 f \beta ^2 \phi ^5 \lambda' r^4+\rho \alpha ^2  (-64 f \pi   (r \lambda'+4 )^2 r^5+128 f \pi  \beta  \phi  \lambda'  (r \lambda'+4 )  (f  (r \lambda'+16 )-4 ) r^4 \nn\\
&\quad&-1024 f (3 f-1) \pi  \beta ^2 \phi ^2  (\lambda' )^2  (f  (r \lambda'+7 )-1 ) r^3+2048 f^2 (1-3 f)^2 \pi  \beta ^3 \phi ^3  (\lambda' )^3 r^2 ) \nn\\
&\quad&+\alpha   (-f \phi  (3 \phi +2)  (r^2  (\lambda' )^2+2 r \lambda'-8 ) r^5-4 \beta  \phi ^2  ( (8 r^2  (\lambda' )^2+4 r \lambda' \nn\\
&\quad&+\phi   (3 r^2  (\lambda' )^2+8 r \lambda'-40 )-40 ) f^2+ (4  (r^2  (\lambda' )^2+r \lambda'+12 )+\phi   (5 r^2  (\lambda' )^2 \nn\\
&\quad&-4 r \lambda'+48 ) ) f-8 (\phi +1) ) r^3-16 f \beta ^2 \phi ^3 \lambda'  ( (-6 r \lambda'+\phi   (4 r^2  (\lambda' )^2+r \lambda'-6 )+12 ) f^2 \nn\\
&\quad&+ (4 r \lambda'+\phi   (6 r \lambda'+20 )-8 ) f+2 r \lambda'+\phi   (r \lambda'-14 )-4 ) r^2+64 f \beta ^3 \phi ^5  (\lambda' )^2  ( (6 r \lambda'-9 ) f^3 \nn\\
&\quad&+ (15-4 r \lambda' ) f^2-7 f+1 ) r )+\alpha ^2  (24 f (f+1) (\phi +1) \lambda'  (r \lambda'+4 ) r^4 \nn\\
&\quad&-16 \beta   ( ( (3 r^4  (\lambda' )^4+22 r^3  (\lambda' )^3+36 r^2  (\lambda' )^2-84 r \lambda'-56 ) \phi ^2+ (3 r^4  (\lambda' )^4+26 r^3  (\lambda' )^3 \nn\\
&\quad&+36 r^2  (\lambda' )^2-96 r \lambda'-64 ) \phi +4  (r \lambda'-2 )  (r \lambda'+1 )^2 ) f^3+24 (\phi +1)  (r \lambda'+ \nn\\
&\quad&\phi   (r^2  (\lambda' )^2+5 r \lambda'+5 )+1 ) f^2-12 (\phi +1)  (r \lambda'+\phi   (r^2  (\lambda' )^2+3 r \lambda'+6 )+2 ) f \nn\\
&\quad&+8 (\phi +1)^2 ) r-768 (f-1)^2 f^3 \beta ^3 \phi ^3 (\phi +1)  (\lambda' )^3  (r \lambda'-2 ) \nn\\
&\quad&+384 f \beta ^2 \phi ^2 (\phi +1) \lambda'  ( (r^3  (\lambda' )^3+4 r^2  (\lambda' )^2-7 r \lambda'-4 ) f^3+ (-r^3  (\lambda' )^3-2 r^2  (\lambda' )^2 \nn\\
&\quad&+11 r \lambda'+8 ) f^2- (5 r \lambda'+4 ) f+r \lambda' ) )+P(\rho )  ( (-384 f \pi   (r \lambda'+4 ) r^5 \nn\\
&\quad&+128 \pi  \beta   ( (8  (r^2  (\lambda' )^2-r \lambda'-2 )+\phi   (r^3  (\lambda' )^3+18 r^2  (\lambda' )^2+64 r \lambda'-16 ) ) f^2 \nn\\
&\quad&+ (8  (r \lambda'+4 )-2 \phi   (r^2  (\lambda' )^2+12 r \lambda'-16 ) ) f-16 (\phi +1) ) r^3 \nn\\
&\quad&+1024 f \pi  \beta ^2 \phi ^2 \lambda'  ( (r^2  (\lambda' )^2-14 r \lambda'-18 ) f^2+ (r^2  (\lambda' )^2+12 r \lambda'+28 ) f \nn\\
&\quad&-2  (r \lambda'+5 ) ) r^2-2048 (f-1) f \pi  \beta ^3 \phi ^3  (\lambda' )^2  (3  (r \lambda'-6 ) f^2+ (r \lambda'+12 ) f-2 ) r ) \alpha ^2 \nn\\
&\quad&+ (128 \pi  \beta  \phi ^2  (3 f  (r \lambda'-2 )+2 ) r^5+1024 f \pi  \beta ^2 \phi ^3 \lambda' r^4 ) \alpha  )\Big)\Big/
\Big(4 f^2 \alpha   (4 (f-1) \beta  \phi -r^2 )  \nn\\
&\quad& (-4 (3 f+1) r^2 \beta ^2 \lambda' \phi ^4+\beta   (8 r^3+96 (f-1) f \alpha  \beta   (\lambda' )^2 r- (r^4+192 f^2 \alpha  \beta \nn\\
&\quad& -192 f \alpha  \beta  ) \lambda' ) \phi ^3-8 \alpha  \beta   (-12 \beta  \lambda'  (r \lambda'-2 ) f^2+3  ( (r^3+4 \beta  r )  (\lambda' )^2 \nn\\
&\quad&+ (-32 \pi  \beta  P(\rho ) r^2+3 r^2-8 \beta  ) \lambda'-4 r ) f+r  (12-r (32 \pi  \beta  P(\rho )+3) \lambda' ) ) \phi ^2 \nn\\
&\quad&+2 r \alpha   (32 \pi  \beta  P(\rho )  (r \lambda'-8 ) r^2+12  (r^2-4 \beta  )+3  (r^3+4 \beta  r ) \lambda' \nn\\
&\quad&-12 f \beta   (r^2  (\lambda' )^2+3 r \lambda'-4 ) ) \phi +6 r^3 \alpha   (r \lambda'+4 ) )\Big) \,,
\eea
\bea
f' &=& F_2 (r, \lambda^{\prime}, f^{\prime}, \phi, \rho) \nn\\
&\equiv& \Big(4096 \pi ^2 \alpha ^2 \beta  P(\rho )^2  (f  (r \lambda'-2 )+2 ) r^5+\beta  \phi ^4  (3 f  (r \lambda'+2 )+2 ) r^5 \nn\\
&\quad& +8 f (1-3 f) \beta ^2 \phi ^5 \lambda' r^4+\rho \alpha ^2  (64 f \pi   (r \lambda'+4 )^2 r^5-128 f \pi  \beta  \phi  \lambda'  (r \lambda'+4 )  \nn\\
&\quad& (f  (r \lambda'+16 )-4 ) r^4+1024 f (3 f-1) \pi  \beta ^2 \phi ^2  (\lambda' )^2  (f  (r \lambda'+7 )-1 ) r^3 \nn\\
&\quad&-2048 f^2 (1-3 f)^2 \pi  \beta ^3 \phi ^3  (\lambda' )^3 r^2 )+\alpha ^2  (12 f (\phi +1)  (r \lambda'+4 )  \nn\\
&\quad& (f  (r^2  (\lambda' )^2+2 r \lambda'+4 )-2  (r \lambda'+2 ) ) r^3-192 f (3 f-1) \beta ^2 \phi ^2 (\phi +1)  (\lambda' )^2  \nn\\
&\quad& ( (r \lambda'-2 ) f^2+ (r \lambda'+4 ) f-2 ) r-32 \beta   ( (-2 r^3  (\lambda' )^3+6 r \lambda'+\phi ^2  (r^3  (\lambda' )^3 \nn\\
&\quad&+6 r^2  (\lambda' )^2+36 r \lambda'+4 )+\phi   (-r^3  (\lambda' )^3+6 r^2  (\lambda' )^2+42 r \lambda'+8 )+4 ) f^3 \nn\\
&\quad&-3  ( (r^3  (\lambda' )^3+8 r^2  (\lambda' )^2+20 r \lambda'+4 ) \phi ^2+ (r^3  (\lambda' )^3+8 r^2  (\lambda' )^2+24 r \lambda'+8 ) \phi  \nn\\
&\quad&+4 r \lambda'+4 ) f^2+6 (\phi +1)  (r \lambda'+\phi   (r^2  (\lambda' )^2+4 r \lambda'+2 )+2 ) f-4 (\phi +1)^2 ) r ) \nn\\
&\quad&+\alpha   (f \phi   (r \lambda'+4 )  (2 r \lambda'+3 \phi   (r \lambda'+2 )+8 ) r^5-2 \beta  \phi ^2  ( (8  (r^2  (\lambda' )^2+8 r \lambda'-2 ) \nn\\
&\quad&+\phi   (r^3  (\lambda' )^3+14 r^2  (\lambda' )^2+28 r \lambda'-48 ) ) f^2-2  (4 r \lambda'  (r \lambda'+4 )  \nn\\
&\quad&+\phi   (5 r^2  (\lambda' )^2+10 r \lambda'-16 ) ) f+16 (\phi +1) ) r^3+16 f (3 f-1) \beta ^2 \phi ^3 \lambda'  (-2 r \lambda' \nn\\
&\quad&+\phi   (12-r \lambda' )+f  (2 r \lambda'+\phi   (r^2  (\lambda' )^2+r \lambda'-12 )-4 )+4 ) r^2 \nn\\
&\quad&-32 f (1-3 f)^2 \beta ^3 \phi ^5  (\lambda' )^2  (f  (r \lambda'-2 )+2 ) r )+ \nn\\
&\quad&P(\rho )  ( (384 f \pi   (r \lambda'+4 ) r^5-256 \pi  \beta   ( (4 r^2  (\lambda' )^2-4 r \lambda'+\phi   (11 r^2  (\lambda' )^2 \nn\\
&\quad&+46 r \lambda'+8 )-8 ) f^2+ (-r^2 \phi   (\lambda' )^2+2 r (2-5 \phi ) \lambda'+16 ) f-8 (\phi +1) ) r^3 \nn\\
&\quad&+2048 f (3 f-1) \pi  \beta ^2 \phi ^2 \lambda'  (-r \lambda'+4 f  (r \lambda'+1 )-4 ) r^2 \nn\\
&\quad&-4096 (f-1) f (1-3 f)^2 \pi  \beta ^3 \phi ^3  (\lambda' )^2 r ) \alpha ^2+ (512 f (3 f-1) \pi  r^4 \beta ^2 \phi ^3 \lambda' \nn\\
&\quad&-256 \pi  r^5 \beta  \phi ^2  (r \lambda' f+f+1 ) ) \alpha  )\Big)\Big/
\Big(2 f \alpha   (-4 r- (r^2+4 (1-3 f) \beta  \phi  ) \lambda' )  \nn\\
&\quad& (-4 (3 f+1) r^2 \beta ^2 \lambda' \phi ^4+\beta   (8 r^3+96 (f-1) f \alpha  \beta   (\lambda' )^2 r- (r^4+192 f^2 \alpha  \beta  \nn\\
&\quad&-192 f \alpha  \beta  ) \lambda' ) \phi ^3-8 \alpha  \beta   (-12 \beta  \lambda'  (r \lambda'-2 ) f^2+3  ( (r^3+4 \beta  r )  (\lambda' )^2 \nn\\
&\quad&+ (-32 \pi  \beta  P(\rho ) r^2+3 r^2-8 \beta  ) \lambda'-4 r ) f+r  (12-r (32 \pi  \beta  P(\rho )+3) \lambda' ) ) \phi ^2 \nn\\
&\quad&+2 r \alpha   (32 \pi  \beta  P(\rho )  (r \lambda'-8 ) r^2+12  (r^2-4 \beta  )+3  (r^3+4 \beta  r ) \lambda'-12 f \beta  \nn\\
&\quad&  (r^2  (\lambda' )^2+3 r \lambda'-4 ) ) \phi +6 r^3 \alpha   (r \lambda'+4 ) )\Big) \,, \\
\phi' &=& F_3 (r, \lambda^{\prime}, f^{\prime}, \phi, \rho) \equiv \frac{8 \alpha  \phi +8 \alpha -8 \alpha  f (\phi +1)  (r \lambda'+1 )+64 \pi  \alpha  r^2 P(\rho )-r^2 \phi ^2}{4 \alpha  f  (\lambda'  (4 \beta  (1-3 f) \phi +r^2 )+4 r )} \,, \\
\rho' &=& F_4 (r, \lambda^{\prime}, \rho) \equiv -\frac{\lambda' (P(\rho )+\rho)}{2 P^{\prime}(\rho )} \,.
\eea
The detailed expressions for $A_1(r)$ and $A_2(r) $ in Eq.~(\ref{firstorder}) are given by
\bea
A_1 &=& -\Big(16 \pi  \alpha  r (P(\rho )+\rho )  (\lambda'  (4 \beta  (1-3 f) \phi +r^2 )+4 r )\Big)
\Big/\Big(f  (\alpha  \phi   (-8 \beta -4 \beta  f  (r \lambda'-2 ) \nn\\
&\quad& -64 \pi  \beta  r^2 P(\rho )+r^3 \lambda'+4 r^2+4 \beta  r \lambda' )-4 \alpha  \beta  (f-1) \phi ^2  (r \lambda'-2 ) \nn\\
&\quad& +\beta  r^2 \phi ^3+\alpha  r^2  (r \lambda'+4 ) )\Big) \,, \\
A_2 &=&  \Big(-8 f^2 r \alpha  \beta   (4 r+ (r^2+4 (1-3 f) \beta  \phi  ) \lambda' )^2  (\phi ' )^2-2 f  (-2 \beta  \phi ^2  (4 r^3 \nn\\
&\quad& +8 f (1-3 f) \alpha  \beta   (\lambda' )^3 r^2+8 (3 f-1) \alpha  \beta   (5 f+3 r f'-1 )  (\lambda' )^2 r \nn\\
&\quad& + (r^4-96 \alpha  \beta  f' r+576 f^2 \alpha  \beta +96 f \alpha  \beta   (3 r f'-2 ) ) \lambda' ) r-2 \alpha  \beta  \phi   (48 \beta  \lambda'  (r \lambda'+1 ) f^2 \nn\\
&\quad& + ( (\lambda' )^3 r^4-80 r-2  (3 r^3+8 \beta  r )  (\lambda' )^2-4  (96 \pi  \beta  P(\rho ) r^2+15 r^2+16 \beta  ) \lambda' ) f \nn\\
&\quad& -3 r^2 f'  (r \lambda'+4 )^2+4  ( (\lambda' )^2 r^3-4 r+ (32 \pi  \beta  P(\rho ) r^2+3 r^2+4 \beta  ) \lambda' ) ) r \nn\\
&\quad& -r^4 \alpha   (r \lambda'+4 )^2+4 (3 f-1) \beta ^2 \phi ^3 \lambda'  (r^3-24 \alpha  \beta  f' \lambda' r-24 f^2 \alpha  \beta  \lambda'  (r \lambda'-6 ) \nn\\
&\quad& +8 f \alpha  \beta  \lambda'  (9 r f'+r \lambda'-6 ) ) ) \phi '-r  (48 \alpha  \beta ^2 \phi ^2 (\phi +1)  (\lambda' )^2  (3 r \lambda'-20 ) f^3 \nn\\
&\quad& +8 \beta  \phi   (6 r \beta  \lambda' \phi ^3-2 \alpha  \beta  \lambda'  (2 \lambda'  (3 r \lambda'-22 )+3 f'  (7 r \lambda'+4 ) ) \phi ^2  \nn\\
&\quad& +\alpha   (-3  (r^3+4 \beta  r )  (\lambda' )^3+2  (48 \pi  \beta  \rho  r^2+48 \pi  \beta  P(\rho ) r^2+7 r^2-21 \beta  f' r+44 \beta  )  (\lambda' )^2 \nn\\
&\quad& -4  (96 \pi  \beta  P(\rho ) r-25 r+6 \beta  f' ) \lambda'+8 ) \phi +\alpha   (-3 r^3  (\lambda' )^3+14 r^2  (\lambda' )^2+100 r \lambda' \nn\\
&\quad& +8 ) ) f^2+ (-16 r \beta ^2  (3 r f'+1 ) \lambda' \phi ^4+8 \beta   (2 r \alpha  \beta   (\lambda' )^3-16 \alpha  \beta   (\lambda' )^2 \nn\\
&\quad& +12 \alpha  \beta  f'  (r \lambda'+4 ) \lambda'-r^2 ) \phi ^3-8 \alpha  \beta   (- (r^3+2 \beta  r )  (\lambda' )^3+ (32 \pi  \beta  \rho  r^2 \nn\\
&\quad& +32 \pi  \beta  P(\rho ) r^2- (r^2+4 \beta  )  (3 r f'-4 ) )  (\lambda' )^2-4  (32 \pi  \beta  P(\rho )  (3 r f'+1 ) r-9 r \nn\\
&\quad& +3  (r^2+4 \beta  ) f' ) \lambda'+8 ) \phi ^2+\alpha   ( (r^5+8 \beta  r^3 )  (\lambda' )^3-8 r^2 \beta   (8 \pi  \rho  r^2-3 f' r+4 )  (\lambda' )^2 \nn\\
&\quad& -16 r  (16 \pi  \beta  \rho  r^2+3 r^2-6 \beta  f' r+18 \beta  ) \lambda'-64  (2 r^2+\beta  )-64 \pi  r^2 \beta  P(\rho )  (r^2  (\lambda' )^2 \nn\\
&\quad& +4 r \lambda'-8 ) ) \phi +r^2 \alpha   (r \lambda'-8 )  (r \lambda'+4 )^2 ) f-f'  (4 r+ (r^2+4 \beta  \phi  ) \lambda' )  \nn\\
&\quad&  (-2 r^2 \beta  \phi ^3+4 \alpha  \beta   (r \lambda'+4 ) \phi ^2+\alpha   (128 \pi  \beta  P(\rho ) r^2+ (r^2+4 \beta  )  (r \lambda'+4 ) ) \phi \nn\\
&\quad&  +r^2 \alpha   (r \lambda'+4 ) ) )-2 f r \beta  \phi   (-96 \alpha  \beta  \phi  (\phi +1) f^2-4  (3 r^2 \beta  \phi ^3-32 \alpha  \beta  \phi ^2+6 r^2 \alpha  \phi  \nn\\
&\quad& -32 \alpha  \beta  \phi +6 r^2 \alpha  ) f+ (r^2+4 \beta  \phi  )  (r^2 \phi ^2-8 \alpha  \phi -8 \alpha  ) \nn\\
&\quad& -64 \pi  r^2 \alpha   (r^2+4 (1-3 f) \beta  \phi  ) P(\rho ) ) \lambda''\Big)\Big/
\Big(2 f r \alpha   (4 r \nn\\
&\quad& + (r^2+4 (1-3 f) \beta  \phi  ) \lambda' )^2  (r+\phi   (r-4 f \beta  \phi ' ) )\Big)\,.
\eea

\section{Analytical expression of SLy EOS} \label{appendixB}

The analytic representation of SLy EOS used in our work is given by~\cite{Haensel:2004nu}
\bea
 \zeta &=& \frac{a_1 +a_2 \xi +a_3 \xi^3}{1 +a_4 \xi} H(a_5 (\xi - a_6)) + (a_7 +a_8 \xi) H(a_9 (a_{10} -\xi )) \nn\\
 &\quad& + (a_{11} +a_{12} \xi) H(a_{13} (a_{14} -\xi ))  + (a_{15} +a_{16} \xi) H(a_{17} (a_{18} -\xi )) \,, \label{slyeos}
\eea
where
\be
\zeta = \log p \,,\quad \xi = \log \rho \,,\quad H(x) = \frac{1}{1+ e^x} \,,
\ee
and the parameters $a_i$ are given in Table~\ref{SLy}.
\begin{table}[htbp]
  \caption{Parameters $a_i$ in Eq.~(\ref{slyeos}).}
  \begin{center}
  \begin{tabular}{ llllllllllll }
  \hline\hline
 $i$ & $a_i$ &  $i$ & $a_i$ &  $i$ & $a_i$ & $i$ & $a_i$& $i$ & $a_i$& $i$ & $a_i$ \\ 
  \hline
  1 $\quad$& 6.22  $\quad$ &2 $\quad$&  6.121  $\quad$  & 3 $\quad$&  0.005925  $\quad$& 4 $\quad$&   0.16326  $\quad$& 5 $\quad$&  6.48  $\quad$& 6$\quad$ &   11.4971  $\quad$\\ 
  7 &  19.105  $\quad$ &8 &   0.8938  $\quad$  & 9 &  6.54  $\quad$& 10 &   11.4950  $\quad$& 11 &  -22.775  $\quad$& 12 &  1.5707  $\quad$\\ 
  13 & 4.3  $\quad$ &14 &  14.08  $\quad$  & 15 &  27.80  $\quad$& 16 &  -1.653  $\quad$& 17 &  1.50  $\quad$& 18 &  14.67  $\quad$\\ 
  \hline\hline
  \end{tabular}  
  \label{SLy}
  \end{center}
\end{table}

\begin{acknowledgments}
We are grateful to Hai-Shan Liu, H. L\"u, and Rui Xu for useful discussions. 
S.L. and H.Y. were supported in part by the National Natural Science Foundation of China (12105098, 12481540179, 12075084,  11690034, 11947216, and 12005059), the Natural Science Foundation of Hunan Province (2022JJ40264), and the innovative research group of Hunan Province under Grant No. 2024JJ1006. 

\end{acknowledgments}


\end{document}